\documentclass{PoS}

\usepackage{amsfonts}
\usepackage{amsmath}
\usepackage{subfigure} 

\graphicspath{{./figs/}}

\title{Recent progress in staggered chiral perturbation theory}

\ShortTitle{Progress in SChPT}

\author{Jon Bailey, \speaker{Weonjong Lee}, and Boram Yoon, \\
  Lattice Gauge Theory Research Center, CTP, and FPRD, \\
  Department of Physics and Astronomy, \\
  Seoul National University, Seoul, 151-747, South Korea \\
  E-mail: \email{wlee@snu.ac.kr}}

\author{Hyung-Jin Kim, \\
  Physics Department, Brookhaven National Laboratory,
  Upton, NY11973, USA \\
  E-mail: \email{windy510@gmail.com}}

\author{SWME Collaboration}

\abstract{ We present a review on recent progress in staggered chiral
  perturbation theory (SChPT). In the last decade, the scope of the
  application of SChPT has been extended beyond the level of
  calibration into the region of prediction with high precision. SChPT
  becomes an essential tool to do the data analysis reliably for
  physical observables calculated using improved staggered
  fermions. Here, we focus on the following examples: pion spectrum,
  pion decay constants, $\varepsilon_K$, and $\pi-\pi$ scattering
  amplitudes. In each subject, we review the recent progress and
  future prospects. 
}

\FullConference{The 7th International Workshop on Chiral Dynamics,\\
		August 6 -10, 2012\\
		Jefferson Lab, Newport News, Virginia, USA}

\begin{document}

\section{Introduction}
\label{sec:intro}
One of the most popular methods to put quarks on the lattice is
the staggered fermion formalism.
Staggered fermions and their improved versions preserve a part of the
full chiral symmetry exactly and have a superior advantage of the
cheapest cost to run on computers.
However, staggered fermions are born with 4 tastes per flavor by
construction.
The disadvantage is that the SU(4) taste symmetry is broken at finite
lattice spacing ($a>0$), which is recovered only in the continuum
limit ($a=0$).
In this paper, we review recent progress in staggered chiral
perturbation theory (SChPT) which is designed to describe the chiral
property of the physical observables calculated using improved
staggered fermions.

\section{Staggered Chiral Perturbation Theory}
\label{sec:schpt}
It is always possible to expand the lattice fermion action in powers
of the lattice spacing $a$, which is quite useful to improve the
action using the Symanzik program.
In this expansion, the leading terms are the continuum QCD action, and
the higher order terms contain the effect of lattice artifacts.
SChPT is a kind of chiral perturbation theory to describe the
chiral properties of physical observables calculated using the
improved staggered fermions \cite{Lee:1999zxa}.
Hence, we map the lattice QCD action into a series in powers of $a^2$.
The leading terms are the same as the continuum chiral perturbation
theory.
And the $a^2$ terms can map into the effective potential terms, which
are of the same order as the leading terms in the power counting rules
of $p^2 \approx m_\pi^2 \approx m_q \approx a^2 $.
Here, note that part of the power counting rules come directly from
the numerical calculation of the pion multiplet spectrum on the
lattice as in Refs.~\cite{Bae:2008qe,Aubin:2004fs}.
Hence, effectively we may say that SChPT corresponds to a dual
expansion in $p^2 \approx m_\pi^2 \approx m_q$ and $a^2$.
Since it is always possible to determine the complete set of
operators at any given order in $a^2$ directly from the lattice
symmetry group, we can say that SChPT incorporates all the
tastes symmetry breaking effects into a dual series expansion
order by order.
As an accidental byproduct, we proved that the pion spectrum
respects the SO(4) taste symmetry out of the full SU(4) symmetry
at the leading order \cite{Lee:1999zxa,Aubin:2003mg},
which is often called \emph{partial symmetry restoration}.
It turns out that the accidental SO(4) symmetry is a quite good
approximation of the pion multiplet spectrum calculated using improved
staggered fermions, as shown in Ref.~\cite{Bae:2008qe,Aubin:2004fs}.

The continuum QCD physics does not have taste degrees of freedom at all.
Hence, for each sea quark loop, we need to reduce the number of tastes
down to 1.
An efficient method to do this job is the rooting prescription.
At finite lattice spacing where the taste symmetry is broken, the
rooting leads to a non-trivial unphysical effect \cite{Sharpe:2006re}.
Here, we assume that this effect is taken care of properly in
SChPT through the replica trick as suggested in
Ref.~\cite{Bernard:2006zw}, which is strongly supported by the
numerical results in Ref.~\cite{Bazavov:2009bb}.

The SChPT Lagrangian
%
\begin{align}
  \mathcal{L}_\mathrm{LO} =
  &\frac{f^2}{8} \textrm{Tr}(\partial_{\mu}\Sigma \partial_{\mu}
  \Sigma^{\dagger}) - 
  \frac{1}{4}\mu f^2 \textrm{Tr}(M\Sigma+M\Sigma^{\dagger}) 
  + \frac{2m_0^2}{3}(U_I + D_I + S_I)^2 + a^2 \mathcal{V}
\end{align}
where $M$ is a quark mass matrix $\text{diag}(\,m_u,\,m_d,\,m_s)
\otimes \xi_I$, and $\mathcal{V}$ is the effective potential which
reflects the effect of taste symmetry breaking at $\mathcal{O}(a^2)$.
This form of the SChPT Lagrangian was first proposed in
Ref.~\cite{Lee:1999zxa} for a single flavor, later extended to
multiple flavors in Ref.~\cite{Aubin:2003mg}, and further extended to
the next-to-leading order (NLO) in Ref.~\cite{Sharpe:2004is}.

\section{Application of SChPT}
\label{sec:app}
During the last decade, SChPT has become a standard tool to obtain the
fitting functions for data analysis of lattice calculations done using
improved staggered fermions.
It has been applied to the pion spectrum, pion decay constants, $B_K$
and its BSM (beyond the standard model) corrections, decay constants
of the heavy-light mesons, semileptonic form factors of the
heavy-light mesons, and so on.
Hence, it becomes impossible to cover the ingredients of interest
over the whole subject in a few pages.
Instead, we will select a few topics and focus on them to review the
recent progress.
Meanwhile, we also present the future prospects in these subjects.

\subsection{Pion masses and quark masses}
\label{ssec:pion-mass}
Pions made of staggered quarks can have flavor and taste structure
$T_f \otimes \xi_t $, where $T_f$ will be Gell-Mann matrices in the
case of SU(3) flavor symmetry and $\xi_t$ are the generators of the
SU(4) taste symmetry.
For simplicity of explanation, we choose the $\pi^+$ flavor to fix 
the $T_f$ matrix.
Then for a given $\pi^+$, we have 16 choices of $\xi_t$: one ($\xi_I$)
belongs to a singlet irrep and the remaining 15 belong to an adjoint
irrep of SU(4).
The Goldstone pion corresponding to an exactly conserved axial current
has the taste of $\xi_5 \in \mathbf{15}$ adjoint irrep.
The rest of the pions with taste not equal to $\xi_5$ are called
non-Goldstone pions because the corresponding axial current is
not conserved at finite lattice spacing.

At the leading order (LO), it was proved that the pion multiplet
spectrum respects the SO(4) taste symmetry
\cite{Lee:1999zxa,Aubin:2003mg}, which turns out to be a very good
symmetry.
This indicates that there are 5 irreps in the pion multiplet spectrum.
At NLO, it was proved that the SO(4) symmetry is broken down to
$SW_4$\footnote{$SW_4$ is a finite subgroup of the $SO(4)$ group.  The
  $SW_4$ is in the diagonal of the direct product of the taste and
  Euclidean SO(4) symmetries.}  \cite{Lee:1999zxa,Sharpe:2004is}.
Hence, we have 8 irreps for the pion multiplet spectrum at NLO.

For the Goldstone pions, the NLO chiral logs were obtained in
Ref.~\cite{Aubin:2003mg}.
This calculation was extended to the non-Goldstone pions in
Ref.~\cite{SWME:2011aa}.
At present, the SWME collaboration works on the mission to extend the
NLO calculation to mixed actions such as HYP staggered valence quarks
with asqtad sea quarks.

\subsection{Pion decay constants}
\label{ssec:f_pi}
%
%
%
Calculation of pion decay constants on the lattice has multiple
targets.
The first is that we can determine $f_\pi$, $f_K$, $f_K/f_\pi$ ratio,
which lead to a precise determination of $V_{us}$.
The second is that, combined with the pion spectrum, it can determine
Gasser-Leutwyler low energy constants and chiral condensates.

There has been an attempt to explore the non-Goldstone
pion sectors using the pion decay constants \cite{Aoki:1999av}.
Recent progress in lattice calculation of pion decay constants are
mainly focused on the Goldstone pion sector \cite{Bazavov:2009bb}.
The NLO corrections in SChPT were obtained in Ref.~\cite{Aubin:2003uc}.
This work has been recently extended to the non-Goldstone pion sectors
in Ref.~\cite{Bailey:2012jy}.
The SWME collaboration plans to extend this further to the mixed
action case.

\subsection{Indirect CP Violation and $B_K$}
\label{ssec:B_K}
The indirect CP violation observed in the neutral kaon system in
nature is parametrized by $\varepsilon_K$, which is measured in
experiment with extremely high precision as follows,
\begin{equation}
\varepsilon_K = (2.228 \pm 0.011) \times 10^{-3} 
\times e^{i\phi_\varepsilon} \,,
\qquad \phi_\varepsilon = (43.52 \pm 0.05){}^{\circ}\,.
\end{equation}
It is also possible to calculate $\varepsilon_K$ directly from the
standard model.
In the standard model, $\varepsilon_K$ can be expressed in terms of
$B_K$ and $V_{cb}$ \cite{Jang:2012ft}.
Here, $B_K$ is a highly non-perturbative parameter which can be
determined reliably only using lattice QCD.
In the case of $V_{cb}$, there are two independent methods to obtain
from the experiment: one is the exclusive decay channel and the
other is the inclusive decay channel.
The exclusive method is simple and heavily relies on lattice QCD,
whereas the inclusive method is very complicated and relies highly on
QCD sum rules.

In Ref.~\cite{VandeWater:2005uq}, the NLO corrections to $B_K$ are
obtained using the SU(3) SChPT. 
Here, they assumed that the external kaons have the same taste as the
Goldstone pions of the staggered fermion formalism, which simplifies
the technical aspects of calculation a lot.
This work has been extended to the mixed action case and the SU(2)
case in Ref.~\cite{Bae:2010ki}.
In addition, it turns out that the mixed pion (composed of valence
quark and sea antiquark) contribution to $B_K$ completely cancels off
between the numerator and the denominator at NLO, which makes $B_K$ a
gold-plated observable in lattice QCD with a mixed action
\cite{Bae:2010ki}.

Results of the SU(2) SChPT in Ref.~\cite{Bae:2010ki} provide the
fitting functional form which is used for the lattice data analysis
for $B_K$.
Recent results for $B_K$ \cite{Bae:2011ff,Colangelo:2010et} indicate
that there exists a substantial gap of $\approx 3 \sigma$ between the
standard model prediction from lattice QCD and the experimental value
of $\varepsilon_K$ in the exclusive $V_{cb}$ channel
\cite{Jang:2012ft}.
This gap may soon become a probe to identify physics beyond the
standard model (BSM) by constraining models of new physics.
%
%
There are, in general, 4 additional four-fermion operators which can
come from the BSM physics.
These four operators can be parametrized into four B-parameters: $B_i$
($i=2,3,4,5$).
Recently, the chiral behavior of the BSM B-parameters is presented at
NLO in SChPT \cite{Bailey:2012wb}.
This work provides the fitting functional form which is used for the
data analysis for the BSM B-parameters in Ref.~\cite{Bailey:2012bh}.

\subsection{$\pi-\pi$ scattering and phase shift}
\label{ssec:phase-shift}
Let us consider $\pi-\pi$ scattering in the staggered fermion formalism.
By construction, there are five non-degenerate channels of two pion
states in a singlet irrep of the SU(4) taste symmetry instead of a
single two pion state as in the case of Wilson-like fermions.
In other words, the five two pion states are
\begin{align*}
\pi(P)-\pi(P)\,, \quad \pi(A)-\pi(A)\,, \quad \pi(T)-\pi(T)\,,
\quad \pi(V)-\pi(V) \,, \quad \pi(S)-\pi(S) 
\end{align*}
Hence, we have to consider at least $5 \times 5$ S-matrix in staggered
fermion formalism to calculate the $\pi-\pi$ scattering phase shift
in the taste singlet channel.
Recently, Hansen and Sharpe provided a prescription which allows us to
handle the multi-channel scattering and decay problems in
Ref.~\cite{Hansen:2012tf}.
In this prescription, they assume that the multi-channel S-matrix is
unitary.

In the staggered fermion formalism, we use the rooting method in the
sea quark loops in vacuum polarization diagrams.
Since the SU(4) taste symmetry is broken at finite lattice spacing, we
know that the rooting makes it non-unitary to calculate the $\pi-\pi$
scattering diagrams using staggered fermions.
This is bad news in itself.
However, the good news is that we can calculate those terms which
violate unitarity of the S-matrix order by order in SChPT.
The strategy is that we fit the data to the functional form derived
in SChPT and remove the unwanted part of unitarity violation terms by
hand, and we use the remaining part, which corresponds to the unitary
S-matrix.
Then we can apply the Hansen-Sharpe formula to obtain the scattering
phase shifts.

At present, the SWME collaboration is working on the first stage of
how to dissect the unitarity violation part from the $\pi-\pi$
scattering amplitude.

\begin{table}[tbhp]
\begin{center}
\begin{tabular}{c | c | c | c | c }
\hline
\hline
physics   & Goldstone  & Non-Goldstone & mixed         & numerical \\
\hline
\hline
$m_\pi^2$ & $\checkmark$ & $\checkmark$ & $\bigtriangleup$ & $\checkmark$ \\
\hline
$f_\pi$   & $\checkmark$ & $\checkmark$ & $\bigtriangleup$ & $\checkmark$ \\
\hline
$B_K$     & $\checkmark$ & $\times$   & $\checkmark$       & $\checkmark$ \\
\hline
$\pi-\pi$ & \multicolumn{2}{c|}{$\bigtriangleup$}
                                    & $\times$         & $\bigtriangleup$ \\
\hline
$K\rightarrow\pi\pi$
          & \multicolumn{2}{c|}{$\times$}
                                    & $\times$         & $\times$ \\
\hline
$V_{cb}$  & \multicolumn{2}{c|}{$\checkmark$}
                                    & $\times$         & $\checkmark$ \\
\hline
\hline
\end{tabular}
\end{center}
\caption{Current status of SChPT application. See text for key words.}
\label{tab:status}
\end{table}

\section{Summary and conclusion}
In Table \ref{tab:status}, we present the current status in the
application of SChPT to various physics subjects.
Here, the $\checkmark$ symbol represents that the mission is completed,
and $\times$ represents that no work has yet been done.
The $\bigtriangleup$ symbol indicates that we are working on this subject
now, and it will be done in near future.
%


\acknowledgments
We are grateful to Claude Bernard, Maarten Golterman, and Stephen
Sharpe for private communications.
The research of W.~Lee is supported by the Creative Research
Initiatives Program (2012-0000241) of the NRF grant funded by the
Korean government (MEST).
W.~Lee would like to acknowledge the support from KISTI supercomputing
center through the strategic support program for the supercomputing
application research [No. KSC-2011-G2-06].
%
%

\bibliographystyle{JHEP}
\bibliography{ref}

\end{document}